\documentstyle[aaspp4]{article}
\begin{document}

\lefthead{Elmegreen, et al.}
\righthead{Dust Spirals in NGC 2207}
\slugcomment{submitted to ApJ Letters}

\title{Dust Spirals and Acoustic Noise in the
Nucleus of the Galaxy NGC 2207}

\author{B. G. Elmegreen\altaffilmark{1},
D. M. Elmegreen\altaffilmark{2}, E. Brinks\altaffilmark{3},
C. Yuan\altaffilmark{4,5}, M. Kaufman\altaffilmark{6},
M. Klari\'c\altaffilmark{7}, L. Montenegro\altaffilmark{5},
C. Struck\altaffilmark{8}, M. Thomasson\altaffilmark{9}
}
\altaffiltext{1}{IBM Research Division, T.J. Watson Research Center,
P.O. Box 218, Yorktown Heights, NY 10598, USA, bge@watson.ibm.com}
\altaffiltext{2}{Department of Physics and Astronomy, Vassar College, 
Poughkeepsie, NY 12604, elmegreen@vaxsar.vassar.edu}
\altaffiltext{3}{Depto. de Astronom\'{\i}a, Universidad de Guanajuato,
Apdo. Postal 144, Guanajuato, Gto. 36000, M\'exico, ebrinks@
coyotl.astro.ugto.mx}
\altaffiltext{4}{Institute of Astronomy and Astrophysics,
Academia Sinica, Nankang, Taipei, Taiwan, ROC, yuan@biaa7.biaa.sinica.edu.tw
}
\altaffiltext{5}{City College of New York, 138th St on Convent Ave., New York, NY 10031}
\altaffiltext{6}{Department of Physics and Department of Astronomy,
Ohio State University, 174 West 18th Avenue, Columbus OH 43210,
rallis@mps.ohio-state.edu}
\altaffiltext{7}{Midlands Technical College, Columbia, SC 29202,
klaricm@mtc.mid.tec.sc.us}
\altaffiltext{8}{Department of Physics and Astronomy, Iowa State University,
Ames, IA 50010, curt@iastate.edu}
\altaffiltext{9}{Onsala Space Observatory, S-439 92 Onsala, Sweden,
magnus@oso.chalmers.se}

\begin{abstract}
Observations with the Hubble Space Telescope reveal an irregular
network of dust spiral arms in the nuclear region of the interacting
disk galaxy NGC 2207. The spirals extend from $\sim50$ pc to $\sim300$ pc 
in galactocentric radius, with a projected width of $\sim20$ pc. Radiative
transfer calculations determine the gas properties of the
spirals and the inner disk, and imply a factor of $\sim4$ local 
gas compression in the spirals.  The gas is not strongly self-gravitating,
nor is there a nuclear bar, so the spirals could not have formed by the
usual mechanisms applied to main galaxy disks. Instead, they
may result from acoustic instabilities that amplify
at small galactic radii. Such instabilities may promote gas accretion
into the nucleus. 
\end{abstract}

Subject headings: hydrodynamics --- instabilities --- 
ISM: structure --- galaxies: individual: NGC 2207 ---
galaxies: nuclei

\section{Introduction}

Recent high resolution observations of the nuclear regions of some
spiral galaxies show irregular spiral patterns in dust and gas
that are not continuations of the main spiral arms or 
driven by obvious bars. 
Examples in the recent literature are 
NGC 278 (Phillips et al. 1996),
M81 (Devereux et al. 1997), 
M87 (Ford et al. 1994; Dopita et al. 1997), 
NGC 4414 (Thornley \& Mundy 1997),
NGC 5188 and NGC 6810 (Carollo, Stiavelli, \& Mack 1998).

This {\it Letter} reports
similar spirals within the inner 500 pc of NGC 2207 
that were found in the course of a larger program to study
this galaxy and its interacting neighbor, IC 2163, with the {\it Hubble
Space Telescope}.
NGC 2207 and IC 2163 were previously observed 
from the ground at optical and radio (21 cm)
wavelengths (Elmegreen et al. 1995a; hereafter Paper I), 
and modeled by computer simulations to
reproduce the peculiar morphologies, internal velocities, and orbital
motions of the main galaxy disks (Elmegreen et al. 1995b).

The nuclear spirals in NGC 2207 are too small to observe by any 
present-day instrument other than
HST, so we derive their properties 
and those of the nuclear disk 
using only the HST data, combined with radiative transfer techniques.   
The disk is found to be too weakly self-gravitating for the spirals to be
density waves. 
We investigate other possible origins for the observed structure. 

\section{Observations}

In May 1996, we observed the nuclear region of NGC 2207 with the
refurbished Planetary Camera of the Hubble Space Telescope at 0.0455
arcsec resolution, using filters at four passbands, U, B, V, and I. A color
image made from the 
V and I bands of a region around the nucleus is shown on
the left in figure 1. The inner HI disk (Paper I) 
is inclined by $35^\circ$ and
has a major axis position angle of $\sim140^\circ$;
approximately the same inclination applies to the nuclear disk seen here
(from the ellipticity of the intensity distribution), but the position
angle is $\sim189^\circ$. (Note the arrow in the bottom--left
of the HST image, which indicates north.) The image on the right in
figure 1 is an unsharp-masked version of the V image, which highlights
the dust structure. The angular scale of the figure
is 160 pixels = $7.28$ arcsec on a vertical side, which is 1.24 kpc for
a distance of 35 Mpc (Paper I).

Several dark dust streamers are evident in the figure. They are peculiar
in several respects: (1) they have no accompanying stellar spiral arms
or star formation knots (confirmed by U and B images, which are not shown
here because the S/N is lower than for the V and I images), as do
dustlanes in the main disks of spiral galaxies, (2) they have much
larger pitch angles ($\sim 50^\circ$) than main-disk spirals (which have
pitch angles of $\sim15^\circ$), (3) there are many of them,
particularly in the outer parts, forming a dark pinwheel with some
intersections, not just a regular 2-arm spiral as in grand-design galaxy
disks, and (4) they are very short, with the shortest only several tens
of parsecs long, and the longest several hundred parsecs.

The nucleus has an exponentially varying light profile near the center,
as shown on the top of figure 2. The curves are intensity scans across
the nucleus in V and I bands, averaged over a scan width of 5 pixels,
along a strip parallel to the two arrows shown at the top of the
left hand image in figure 1.
There is no power-law cusp (even on single pixel-wide scans), and hence
no obvious black hole that might be revealed by such a cusp (Lauer et al. 1992).
This is consistent with the lack of a nuclear radio continuum source
(Condon 1983; Vila et al. 1990)
from an active nucleus. Neither is there a nuclear starburst (Paper I). The
exponential scale length in V and I bands within 0.5 arcsec from the
center is $\sim0.47''$, or 80 pc deprojected. Just beyond this distance,
the deprojected scale lengths change to 360 pc in V and 270 pc in I. 

\section{Radiative Transfer Model}

In order to estimate gas densities, the dust spiral was studied using a
radiative transfer model (Block et al. 1996) 
that was fit to an approximately EW scan
positioned $1.638''=278$ pc north of the nucleus (see arrows, Fig. 1).
The V and I-band intensity profiles along the scan are shown on the
bottom of figure 2, along with the model fits (dashed lines). The model
considers an inclined disk 
with properties that are typical for galaxies:
exponential radial profiles with the same scale lengths  
for the stars, gas and dust; Gaussian profiles perpendicular
to the disk with a smaller scale height $h$ for the gas and dust 
mixture than the 
stars (for which $h=100$ pc is assumed),
and the same scale heights for the dark features as for the
regions between them.  Note that the value of the stellar scale height does not
matter much as long as it is several times larger than the gaseous scale height.
Four dust features with Gaussian
profiles parallel to the disk cause the
absorption dips seen in the scans. 
The fit was made to the I band because
extinction is weakest there; only the ratio of extinction to gas column
density was varied between V and I bands, as appropriate for these two
wavelengths. Scattering is included in the models, but found to be
relatively unimportant. 

Some aspects of the model are uniquely specified by the data. The
observed widths of the dust features have to match the projected
dimensions in the models, and then the midplane densities in the dust
features are determined uniquely by the line-of-sight opacities of the
absorption dips. The intrinsic disk scale length in the model is also
determined by the observations, but only approximately, because internal
absorption flattens the disk profile 
(Byun, Freeman, \& Kylafis 1994).
As a result, the model gives the internal extinction, the 
gas surface density between the dust lanes, and the intrinsic disk
scale length, but each are uncertain by a factor of $\sim2$.

The radiative transfer fit in figure 2 has a V-band extinction
in the midplane at the position of the strip equal to $\sim9$ optical
depths per kiloparsec between the dustlanes. This corresponds to an
average hydrogen density of $\sim6$ cm$^{-3}$ with the conventional
dust-to-gas conversion factor (Bohlin, Savage \& Drake 1978)
and the usual assumption that
gas and dust are well mixed. The perpendicular scale height of the inner
disk is $\sim35$ pc, and the intrinsic exponential scale lengths of the
V and I band emissivities are 170 pc and 220 pc (these are less than the
observed scalelengths because absorption in the disk flattens the
apparent intensity profile). The hydrogen densities in the four
dustlanes, in order from east to west (left to right
in Fig. 2), are 11, 17, 45, and 16
cm$^{-3}$, and their in-plane dimensions are $\sim20$, 20, 5, and 25 pc.
The average ratio of the density in a dustlane to the density between
the dustlanes is $\sim4$. Most of the projected extents of these
dustlanes are from the perpendicular disk scale height. The products of
the line-of-sight dustlane thicknesses and the midplane densities give
effective line-of-sight V-band extinctions through the dust features
of $\sim1.2$, 1.8, 1.2, and 2.1 optical depths, respectively.
The total extinction between the dustlanes, perpendicular to the disk,
is the product of twice the disk scale height and the
opacity, or 0.6 optical depths in V band.

The radiative transfer solution is the only way to estimate the gas density
at high resolution; the radio observations are at much lower resolution.
The radiative transfer suggests that the average hydrogen column density
between the dustlanes at this radius, measured perpendicular to the
disk, is $\sim1.3\times10^{21}$ cm$^{-2}$, which corresponds to a mass
column density of 14 M$_\odot$ pc$^{-2}$ including helium. This is
larger than the inclination-corrected 
VLA\footnote{The
National Radio Astronomy Observatory (NRAO) is operated by Associated
Universities, Inc., under cooperative agreement with the National
Science Foundation.}
measurement (Paper I) of $\sim3$
M$_\odot$ pc$^{-2}$ for HI, including helium, in a $13.5''\times12''$
beam centered on the nucleus.  There are no
direct CO observations of the nucleus, but there are two CO detections
made by one of us (MT) with a $43''$ beam at the Swedish-ESO
Submillimetre Telescope (La Silla, Chile),
centered 1/4-beam E and W of
the nucleus; both detections include the nucleus in the beam pattern.
The integrated CO emissions are each 6.9 K km s$^{-1}$ on
the main beam temperature scale, so the inclination-corrected mass
column density for molecular hydrogen plus helium is 29 M$_\odot$
pc$^{-2}$, assuming the standard conversion factor for CO luminosity to
molecular mass (Strong et al. 1988).  All of these observations
suggest the nuclear disk has a relatively low gas column density,
consistent with the lack of a starburst in the inner region.

\section{Possible Origins for the Nuclear Spirals}

What is the origin of the dust spirals in the nuclear disk of NGC 2207?
Spirals like these are not typically present in galaxies with strong
nuclear or inner ring starbursts, where the dust features are
accompanied by star formation (e.g., NGC 6946 -- Engelbracht et al.
1996),
nor are they as symmetric as dust
features in bars (e.g., NGC 4321 -- Sakamoto et al. 1995; Knapen et al
1995). 
The nuclear dust spirals are also not likely to be
related to the main spiral structures because the
nuclear spirals lie inside the inner Lindblad resonance for the outer
spirals (Elmegreen, Elmegreen \& Montenegro 1992) 
and inside a Q barrier that might come from a bulge.
Both of these features shield the nuclear region from incoming spiral
waves (Bertin et al. 1989). 
The position inside the inner resonance also precludes any
resonance excitation that might come from the outer spirals (Sellwood
\& Lin 1989; Sellwood \& Kahn 1991).

This leads us to suspect that the spirals are formed and dispersed
continuously in the nuclear disk. They could be strong compressions or
shock fronts caused by turbulent motions that are driven by gas
inflow (Struck 1997) 
or other irregularities in the nuclear disk, or they could
result from local instabilities.

The magnetically-driven instability discussed by 
Balbus \& Hawley (1991) is a reasonable choice, discussed already
in this context by Dopita et al. (1997).  However, it 
might not be appropriate for 
low-density nuclear disks like NGC 2207 because it requires 
high angular shear.
The disk in M87 studied by Dopita et al. surrounds a black hole and
probably has the required shear, but the situation in NGC 2207 is
not so clear.  We would prefer 
an explanation for the spirals in NGC 2207 that works with little shear.  

Another well-known instability produces multiple-arm spirals in the main
disks of galaxies, and is driven by the gravity of the stars and gas
(Lin \& Shu 1964; Toomre 1981).
This instability becomes strong when a dimensionless parameter $Q=\kappa
a/(\pi G\sigma)$ has a value of around 2 or less (Toomre 1981), and
it continues to work with small angular shear if $Q\sim1$. 
Here $\kappa$ is
the frequency of weak radial oscillations for slightly displaced
material, $a$ is the rms velocity dispersion, $G$ is the gravitational
constant, and $\sigma$ is the gas mass column density of the disk. We do not
observe $a$ and $\kappa$ in the nuclear region directly because the
angular scale is too small, but we can estimate them from the radiative
transfer solution. 
We consider solid-body rotation at angular rate $\Omega$,
for which $\kappa=2\Omega$, and treat first
a non-self-gravitating disk, where the scale height is
$h=a/\Omega$.
These combine to give $Q=2a^2/(\pi G\sigma h)$.
The radiative transfer solution gives $\sigma\sim14$ M$_\odot$
pc$^{-2}$ and $h=35$ pc in the vicinity of the strip. Thus 
$Q\sim0.3a^2$ with $a$ in km s$^{-1}$.
The VLA data (Paper I) suggest that the entire disk,
including the inner portion, has a relatively large gaseous velocity
dispersion, possibly in excess of 40 km s$^{-1}$. If $a({\rm
km\;s}^{-1})>10$, which is likely, then $Q$ is much larger than 2, 
consistent with our use of $h\sim a/\Omega$. We
note that this estimate for $\Omega\sim a/h\sim0.28$ km s$^{-1}$ pc$^{-1}$
is also consistent with ground-based optical 
observations (Rubin et al. 1983) at $1''$ resolution.
This implies that even the usual expression for $Q$, which is valid
for a self-gravitating disk as well, gives a large value. 
Setting $\Omega>0.1$ km s$^{-1}$ pc$^{-1}$ from Rubin et al., we 
get $Q=\kappa a / \left(\pi G \sigma\right) > 1.1a$ for $a$ in 
km s$^{-1}$.  This is again much larger than 2 for reasonable 
velocity dispersion $a$. 
Because of the large $Q$, 
the self-gravitational instability that drives spirals in the main
disks of galaxies cannot operate in the nuclear disk of NGC 2207.

If self-gravity cannot generate the spirals, then another local 
source must be present.  An intriguing possibility is the natural tendency
for sound waves to amplify at small galactocentric radii
(Montenegro 1998; Montenegro, Yuan \& Elmegreen 1998).
Consider a disk in rigid rotation with
constant surface density (exponential variations in $\sigma$ and $a$ 
and radial variations in $\Omega$
do not affect the results significantly). 
The governing equation for a wave-like perturbation
in column density,
$\sigma_1\sim h_1e^{i\left(\omega t-m\theta\right)}$, is
\begin{eqnarray}
\frac{d^2h_1}{dr^2}+\frac{1}{r}\frac{dh_1}{dr}
+(\frac{{\kappa}^2}{a^2}({\nu}^2-1)-\frac{m^2}{r^2})h_1=0.
\label{bessel}
\nonumber
\end{eqnarray}
Here $\nu\equiv(\omega-m\Omega)/\kappa$, $m$ is the azimuthal wavenumber, 
$r$ is the galactocentric
radius, and
$\Omega$, $\kappa$, and $a$ were defined above.  When
\[
\frac{{\kappa}^2}{a^2}({\nu}^2-1)\equiv{\alpha}^2>0  ,
\]
the solutions are Bessel functions $J_m(\alpha r)$ and $Y_m(\alpha r)$;
$Y_m$ diverges at small $\alpha r$. For ${\alpha}r\sim\kappa r/a$ large,
$J_m$ and $Y_m$ combine to give outgoing and incoming sonic waves. In
NGC 2207, there is little activity in the nucleus; 
the waves generated locally are probably moving in both directions. 
Those that propagate outward become weaker by geometric effects
(amplitude $\propto r^{-1/2}$) and viscous damping. 
Those that propagate inward amplify as $Y$
diverges; then they dissipate by shocks when r approaches the center
near $r=m/{\alpha}\sim ma/\kappa$. At this radius, the 
separation between waves equals the minimum possible distance, which is the
radius of epicyclic motion at the sound speed. Thus waves with high $m$ occupy
mostly the outer regions of the nuclear disk. This result is in
agreement with the observations in figure 1: there are many more dust
features in an annulus at large radius than at small radius.

The waves also propagate in the azimuthal direction with angular speed
${\omega}/m$, as long as ${\nu}^2-1>0$. Waves with different $m$
therefore interact to form complex structures and intersections, as seen
in NGC 2207. The waves may also become sheared by slight differential
rotation to form open trailing spirals, although shear is not required to make
the gas unstable. 

The above solutions assume the frequency $\omega$ is real and the
radial wavenumber, $k\equiv -id{\rm (phase)}/dr$, 
is complex. We can assume alternatively that $k$ is real
and $\omega$ is complex, in which case the waves, fixed in space, grow
in time. The result (Montenegro et al. 1998)
is a growth time for the waves equal to the
propagation time, $r/a$, and because $r/h\sim r\Omega/a\sim6$ in the
part of NGC 2207 considered here, this is also about the orbit time.
After this time, the waves shock and dissipate, as in the solution $Y$
above. 

These properties of sonic noise at small galactocentric radii should be
appropriate for a nuclear galactic disk that is not dominated by a
strong bar, black hole, or starburst. 
They agree well with the characteristics of the
dust features observed at the center of NGC 2207. Acoustic spirals have
not been considered for galaxies before because the main disks
typically have sound speeds much less than orbit speeds, and they also
have much stronger self-gravity. 
Nevertheless, acoustic spirals may be an important agent for driving
gas accretion in some nuclei.

The accretion rate from acoustic spirals 
may be estimated from the energy dissipation rate.
If the waves become mild shocks, then the energy flux
into each shock is $\sim1.5\rho a^3$, and the total shock dissipation rate
is $\sim1.5\rho a^3\times 2h\times mL$ for number of arms $m$ and
arm length $L$.  Dividing this by the orbital energy available, which is 
$\sim0.5\rho v^2 \times 2h \times \pi r^2$ for orbit speed $v$, the
loss rate becomes several times $a^3m/\left(v^2r\right)$,
considering that $L\sim r$.  Now we can use $r\sim ma/\kappa$ to describe
the radius of wave shocking, from the Bessel function solution above,
and then the overall evolution rate is $\sim \kappa\left(a/v\right)^2$.  
The mass accretion time is therefore comparable to the orbit time
multiplied by the square of the ratio of the orbit speed to the sound
speed. 
In nuclear regions like this, 
the orbit speed may be only only a few times the sound speed, 
and then the gas accretion time can be relatively short.
Presumably there is self-regulation, however, because persistent
shocking in acoustic waves might heat the gas to the point where
shocks no longer appear.  Then the accretion process would slow down 
until the gas cools again. 
The overall impact of acoustic waves on accretion is therefore difficult to
assess. 

\section{Summary}

NGC 2207 contains peculiar dust spiral arms in its nuclear
disk that are not obviously related to star formation, bar structure, or
gravity-driven spiral density waves. They could be regions of forced
compression that are stirred, for example, by inflow, but they could also be
amplified acoustic waves, which tend to drive spiral structure
at small relative radii (i.e., $r\kappa/a$ not much larger than 1). 
The resulting spiral arms steepen and shock in an orbit time, 
which is only $\sim2\times10^7$ years in the region studied,
and this shocking ultimately causes the gas to lose angular
momentum and energy.  Then it can accrete to the nucleus. 

Acoustic spirals may be the primary agent for gas accretion in 
the most quiescent nuclear disks, i.e., those with weak
self-gravity and bar forcing, limited access to outer disk
activity, and little shear.  It may also be
an important supplement to other accretion processes, such
as bar-driven shocks 
and the Balbus \& Hawley (1991) instability, in more active regions.
A more quantitative assessment of its role in general nuclear 
environments is difficult to make at the present time.

Support for this work was provided by NASA through grant
number GO-06483-95A from the Space Telescope Science Institute,
which is operated by the Association of Universities for
Research in Astronomy, Inc., under NASA contract NAS5-26555.
Figure 1 was made using the IBM Data Explorer software, with the help
of Dr. Tom Jackman.

\figcaption{
(left) Hubble Space Telescope image of the nuclear region of NGC 2207
in color rgb format, with r taken to be the I-band 
image, b the V-band image, and g the average of these
two images.  
Arrows at the top indicate the strip used to fit the radiative
transfer model.  North is indicated by the arrow in the lower right.       
(right) Unsharp-masked image of the V-band showing intricate
dust structure as dark bands. The unsharp mask was made by subtracting
the convolution of the V image with a Gaussian of 0.5 arcsec 
FWHM from the original V image. }
                                                                                
\figcaption{
(top) Intensity profiles in V and I bands along a strip 5 pixels ($=0.228''$) 
wide through the nucleus, parallel to the 
strip used for the radiative transfer model, 
shown by the arrows in figure 1. 
(bottom) Intensity profiles in V and I bands through a strip north of the
nucleus, and radiative transfer fits (dashed lines) to these profiles. 
}
                                                                                
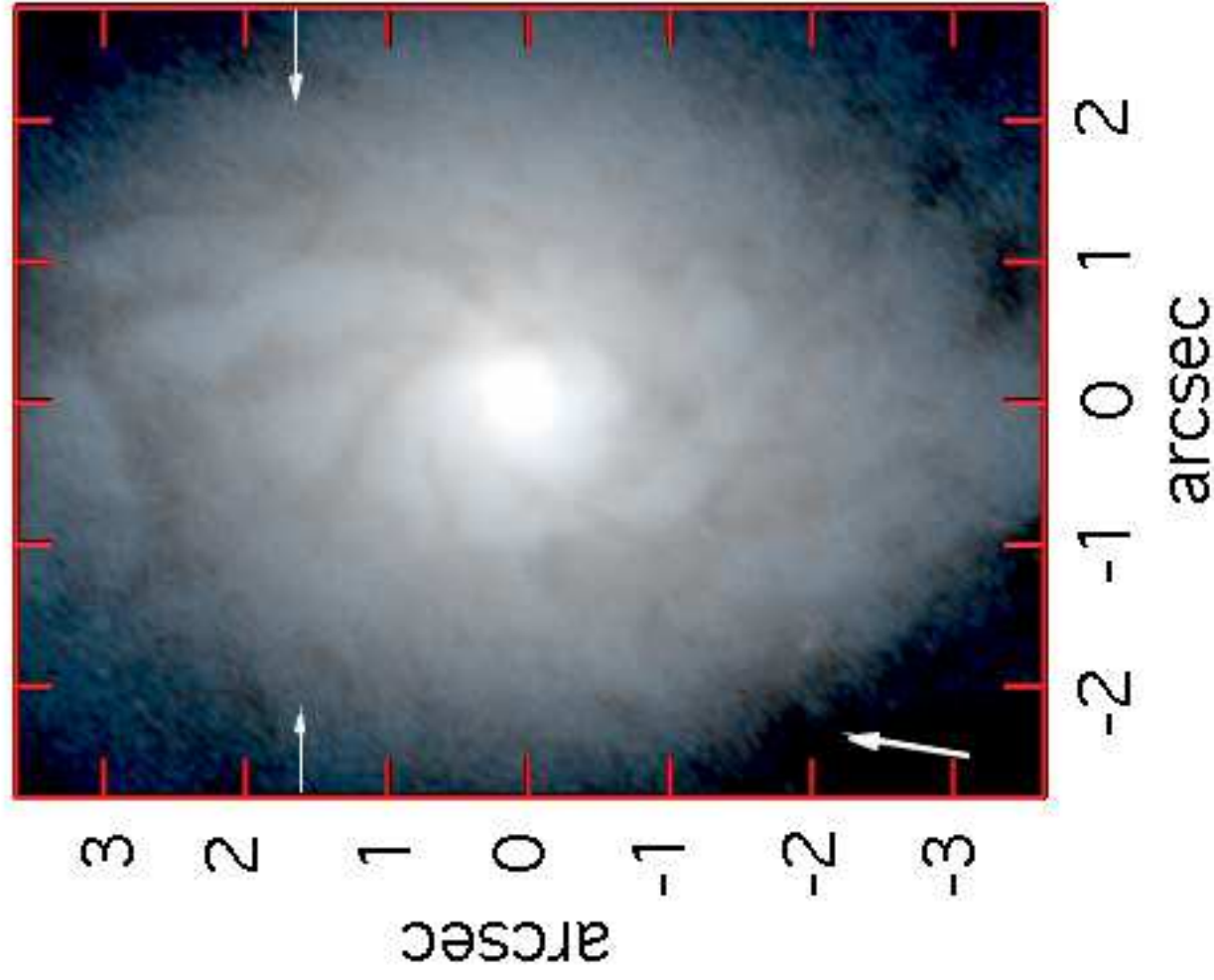
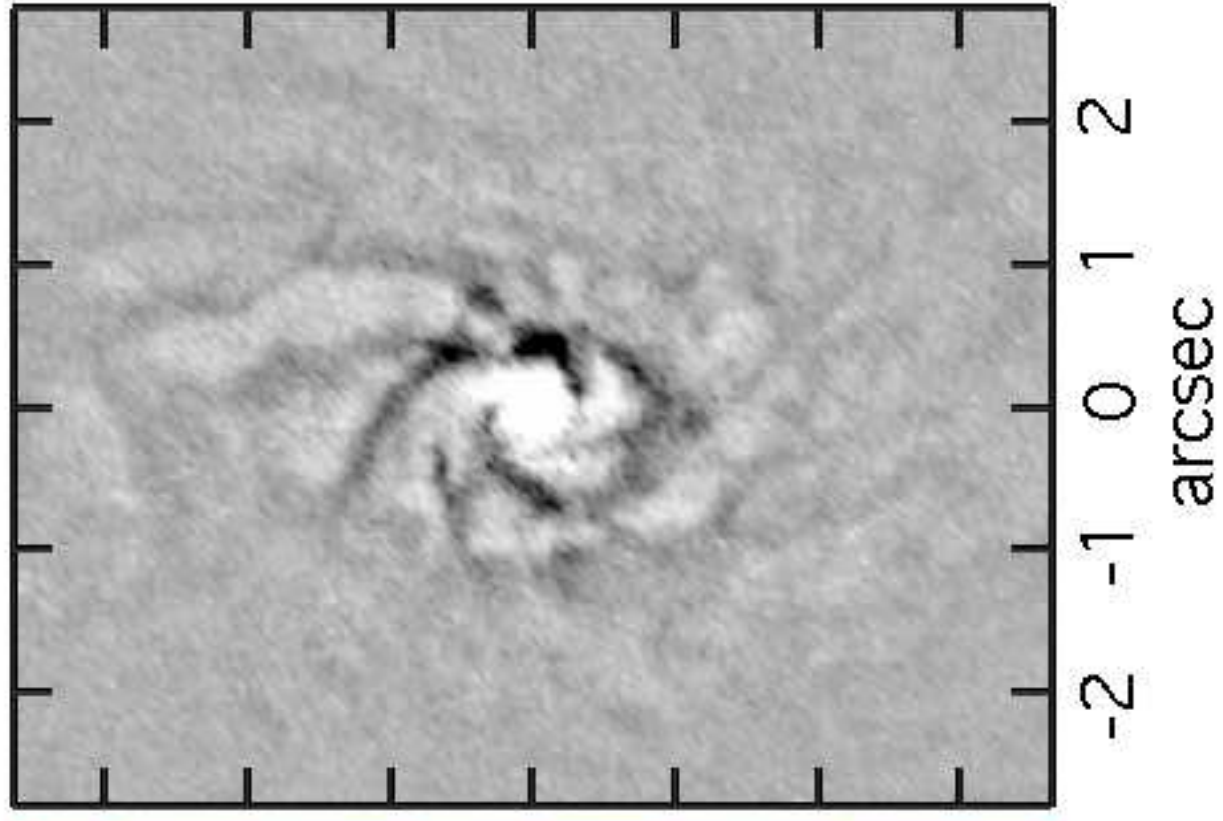
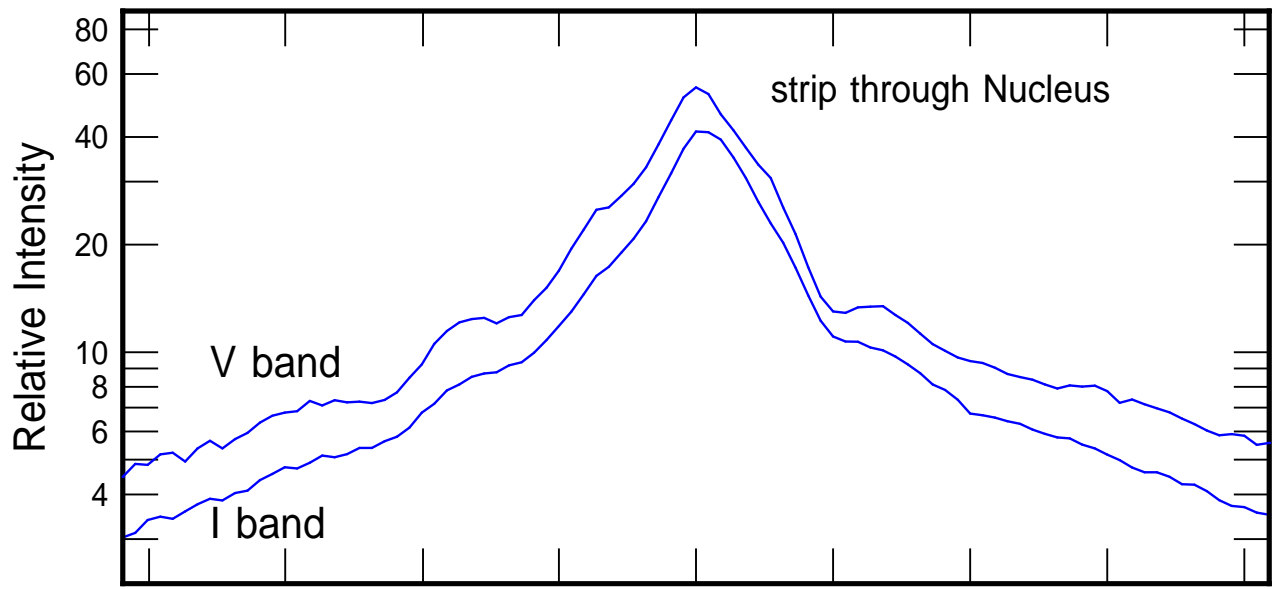
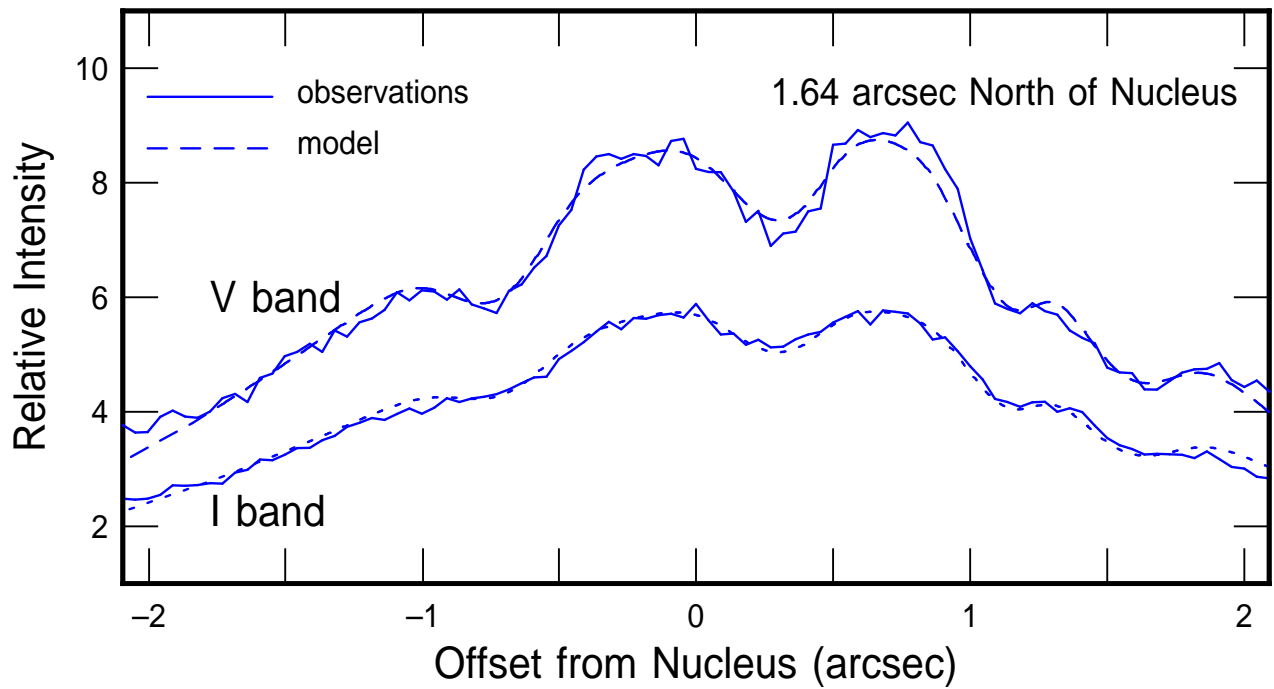
\end{document}